\documentclass[prl,showpacs,twocolumn,superscriptaddress]{revtex4-1}

\usepackage{graphicx}
\usepackage{bm}
\usepackage{amsmath}
\usepackage{amsfonts}

\newcommand{\beq}{\begin{equation}}
\newcommand{\eeq}{\end{equation}}
\newcommand{\beqa}{\begin{eqnarray}}
\newcommand{\eeqa}{\end{eqnarray}}
\newcommand{\vep}{\varepsilon}

\begin{document}


\title{Charge-noise-insensitive gate operations for always-on, exchange-only qubits}

 \author{Yun-Pil Shim}
 \email{ypshim@lps.umd.edu}
 \affiliation{Laboratory for Physical Sciences, College Park, Maryland 20740, USA}
 \affiliation{Department of Physics, University of Maryland, College Park, Maryland 20742, USA}
 \author{Charles Tahan}
 \email{charlie@tahan.com}
 \affiliation{Laboratory for Physical Sciences, College Park, Maryland 20740, USA}
 \date{\today}

\begin{abstract}
We introduce an always-on, exchange-only qubit made up of three localized semiconductor spins that offers a true ``sweet spot" to fluctuations of the quantum dot energy levels. Both single- and two-qubit gate operations can be performed using only exchange pulses while maintaining this sweet spot. We show how to interconvert this qubit to other three-spin encoded qubits as a new resource for quantum computation and communication.
\end{abstract}

\pacs{}
\maketitle



Semiconductor qubits \cite{loss_divincenzo_pra1998,kane_nature1998} are a leading candidate technology for quantum information processing \cite{spin_qubit_review}. Spins can have extremely long quantum coherence due to a decoupling of spin information from charge noise in many materials, and they are small, enabling high density. But these strengths pose a challenge for control as microwave pulses generally result in slow gates with significant potential for crosstalk to nearby qubits. The exchange interaction on the other hand provides a natural and fast method for entangling semiconductor qubits: it can be used to perform two-spin entangling operations with a finite-length voltage pulse or to couple spins with a constant interaction. Exchange also provides a solution to the control problem by allowing a two-level system to be encoded into the greater Hilbert space of multiple physical spins. Following work on decoherence free subspaces and subsystems (DFS) \cite{Duan_Guo_prl1997,Zanardi_Rasetti_prl1997,Lidar_Chuang_Whaley_prl1998}, many multi-spin-based qubits have been proposed and demonstrated with various desirable properties for quantum computing: as examples, 2-DFS (a.k.a. "singlet-triplet") \cite{Levi_prl2002,Petta2005,Maune_Hunter_nature2012,Shulman_Yacoby_science2012}, 3-DFS (a.k.a. "exchange-only") \cite{divincenzo_bacon_nature2000,fong_wandzura_qic2011,Medford_Marcus_nnano2013,Eng_Hunter_sciadv2015}, or 4-DFS qubits \cite{hsieh_whaley_qip2003} of various implementations are possible. The DFS gives some immunity to global field fluctuations, but more importantly it allows for gate operations via a sequence of pair-wise exchange interactions between spins with fast, baseband voltage control on the metallic top-gates, obviating the need for RF pulses. However, charge noise likely limits gate fidelity \cite{Hu_DasSarma_Charge_noise_prl2006,Dial_Yacoby_charge_noise_prl2013} as charge and spin are coupled while spins undergo exchange.

The effects of charge (or other) noise on memory or gate fidelity can be suppressed to a certain extent by taking advantage of natural or engineered ``sweet" spots: a spot in parameter space where critical system properties are minimally effected by certain environmental changes. Sweet spots have been an effective tool to increase the coherence of superconducting qubits \cite{quantronium_Vion_science2002,transmon} and more recently have been applied to exchange-only qubits \cite{RX_theory_Taylor_prl2013,RX_exp_Medford_prl2013,TQD_sweet_spot_Fei_Friesen_prb2015,TQD_sweet_spot_Burkard_prb2015, DQD_sweet_spot_Reed_Hunter_2015}. 
For example, in the ``resonant exchange" (RX) qubit---an encoded qubit made out of 3 quantum dot qubits with ``always-on" exchange interactions and a much higher chemical potential for the middle dot than the outer dots---a partial sweet spot is maintained while microwave control allows for single qubit operations \cite{RX_theory_Taylor_prl2013,RX_exp_Medford_prl2013} ``resonant" with the gap of the 3-spin system. The first derivative of the RX qubit frequency vanishes for one of the two detuning parameters that are affected by charge noise. For two qubit gates, the RX qubit offers a relatively large transition dipole matrix element for two-qubit dipole-dipole coupling, either directly or through a resonator \cite{RX_theory_Taylor_prl2013}. Doherty et al. \cite{RX_2q_gate_Doherty_prl2013} have shown how to perform two-qubit gates between RX qubits with the exchange interaction.   Unfortunately, a full sweet spot for an RX qubit (where the first derivative with respect to both detuning parameters goes to zero) is outside of the (111) singly occupied regime; there, higher order effects limit the coherence of the qubit \cite{TQD_sweet_spot_Burkard_prb2015}. A full sweet spot for a 3-spin qubit was found for a symmetric triple quantum dot (TQD) \cite{TQD_sweet_spot_Fei_Friesen_prb2015}, but in that case one needs to move away from this sweet spot to perform a full set of single-qubit gate operations. 

In this work, we show that there exists a full sweet spot for an exchange-only qubit and we can implement full single- and two-qubit gate operations on this sweet spot with only DC voltage pulses to control the tunneling elements. 
We will call this qubit the "Always-on, Exchange-ONly qubit" (AEON), since both exchange interactions in an AEON qubit are kept on for logical gate operations while remaining on the sweet spot. This is possible since the sweet spot is independent of the tunneling elements. 
The idea of tuning the tunneling barrier directly to control the exchange interaction (as opposed to changing the relative energy-level detuning between two dots), which was the original idea of QD spin qubits \cite{loss_divincenzo_pra1998}, has been successfully demonstrated in recent experiments in Si/SiGe \cite{DQD_sweet_spot_Reed_Hunter_2015} and GaAs/AlGaAs \cite{DQD_symmetric_operation_Martins_Kuemmeth2015} QD devices. The exchange operation via gate-tuning the tunneling barrier near the symmetric operating point (SOP) of the detuning leads to much higher fidelity during the exchange operations than the conventional control of the exchange interaction by tilting (detuning) the double dot system.  
Here we show that a similar approach is possible during simultaneous control of exchange interactions in a three-spin exchange-only qubit. 
This is especially useful for implementing two-qubit logical gates in a single exchange pulse \cite{RX_2q_gate_Doherty_prl2013} which requires the qubits to have exchange interactions on to keep a finite energy gap.  
We also discuss how one can convert an AEON qubit into an RX qubit to take advantage of some of practical advantages of the RX qubit, e.g. initialization/readout and for dipole coupling to resonators. The AEON qubit also smoothly converts to the traditional 3-DFS qubit.


\begin{figure}
  \includegraphics[width=\linewidth]{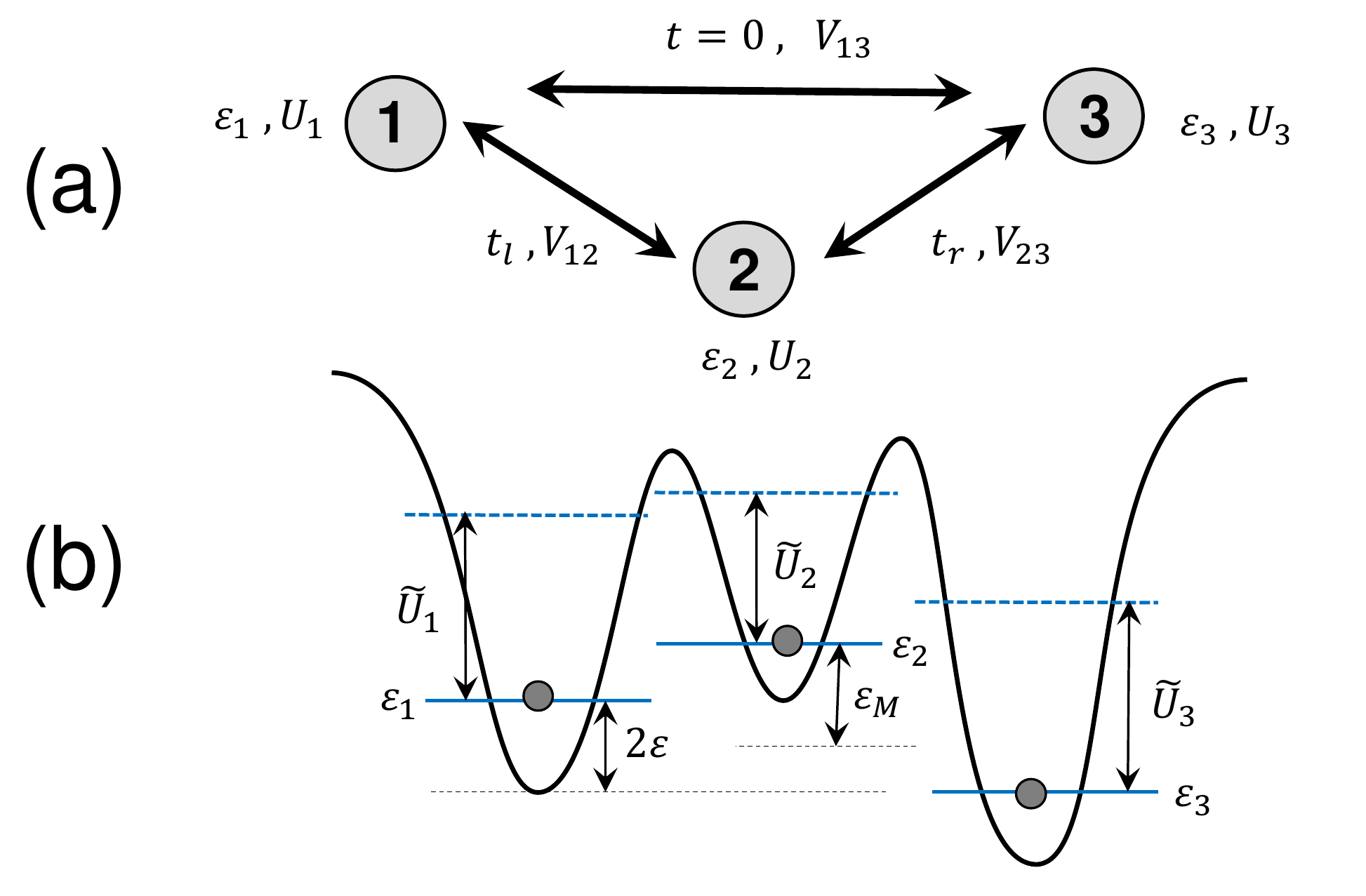}\\
  \caption{(a) Triple quantum dot system with parameters for the Hubbard model. (b) Schematic diagram for the energy levels of the TQD device. Each QD is singly occupied and the dashed blue lines indicate the energy of a second electron if the dot is doubly occupied by a tunneling from a neighboring dot. $\widetilde{U}_i$ is the associated Coulomb energy change due to the double occupancy, including both intra- and inter-dot Coulomb interactions. $\vep$ and $\vep_M$ are the two detuning parameters. 
$\vep$ is the detuning between dot 1 and dot 3, and $\vep_M$ is the detuning of dot 2 with respect to the average of the other two dots. 
  }
  \label{fig:TQD}
\end{figure}

{\it Model--} 
We consider a TQD system as illustrated in Fig. \ref{fig:TQD}, which is described by a Hubbard Hamiltonian:
\begin{eqnarray}\label{eq:Hubbard}
\hat{H} 
&=& \sum_{i=1}^3 \sum_{\sigma} \varepsilon_i c_{i\sigma}^{\dag} c_{i\sigma} 
  + \sum_{i=1}^3 U_i \hat{n}_{i\uparrow} \hat{n}_{i\downarrow} 
  + \frac{1}{2}\sum_{i\neq j} V_{ij} \hat{n}_i \hat{n}_j \nonumber\\
&& + \sum_{<i,j>} \sum_{\sigma} t_{ij} \left( c_{i\sigma}^{\dag} c_{j\sigma} + h.c. \right)
\end{eqnarray}
where $\hat{n}_i$=$\hat{n}_{i\uparrow} + \hat{n}_{i\downarrow}$ and $\hat{n}_{i\sigma}$=$c_{i\sigma}^{\dag} c_{i\sigma}$.
Only linear TQD with $t_{31}$=0 will be considered. We define new parameters for orbital energies of the QDs:
$\bar{\vep}$=$(\vep_1+\vep_2+\vep_3)/3$, $\vep$=$(\vep_1-\vep_3)/2$, $\vep_M$=$\vep_2 - (\vep_1+\vep_3)/2$
and tunneling elements $t_{12}$=$t_l$ and $t_{23}$=$t_r$.
$\bar{\vep}$ just shifts the global energy reference and does not affect the qubit operations. There are two independent detuning parameters that are affected by the charge noise. $\vep$ is the detuning between dot 1 and dot 3, and $\vep_M$ is the detuning of dot 2 with respect to the average of the other two dots [see Fig. \ref{fig:TQD}(b)]. These two detunings affect the qubit coherence.
The tunneling elements $t_l$ and $t_r$ are also susceptible to charge noise, but it is thought that the fluctuations in tunnel couplings are not dominant source of dephasing, compared to the fluctuation in the detunings \cite{Dial_Yacoby_charge_noise_prl2013,DQD_sweet_spot_Reed_Hunter_2015,DQD_symmetric_operation_Martins_Kuemmeth2015}. Therefore, we define sweet spots when the derivative of the qubit energy vanishes with respect to the detunings $\vep$ and $\vep_M$.

The exchange-only qubit consists of three singly (or total spin-1/2) occupied QDs. The exchange interaction arises by virtually occupying one of the dots through tunneling and the relevant charge configurations are (111), (201), (102), (120), (021), (210), (012) where the digits signify the number of electrons in each dot. Two configurations, (210) and (012), are not tunnel coupled to (111) and contribute only higher order corrections and will be neglected in the following. Total spin $\mathbf{S}_{\mathrm{tot}}$ is a good quantum number of the Hubbard Hamiltonian and we focus on the low energy manifold with $S_{\mathrm{tot}}=1/2$. With a uniform external field, Zeeman splitting allows us to further limit the subspace with $S_{\mathrm{tot}}^z=1/2$ for the Zeeman energy much larger than the exchange interaction.   

We choose a set of basis states for charge configurations of (111), (201), (102), (120), and (021):
\begin{eqnarray}
|1\rangle &=& |S\rangle_{13} |\uparrow\rangle_2 
           = \frac{1}{\sqrt{2}} \left[ |\uparrow_1 \uparrow_2 \downarrow_3\rangle -  |\downarrow_1 \uparrow_2 \uparrow_3\rangle  \right]\nonumber\\
|2\rangle &=& -\frac{1}{\sqrt{3}} |T_0\rangle_{13} |\uparrow\rangle_2 + \sqrt{\frac{2}{3}} |T_+\rangle_{13} |\downarrow\rangle_2 \nonumber\\
          &=& -\frac{1}{\sqrt{6}} \left[ |\uparrow_1 \uparrow_2 \downarrow_3\rangle + |\downarrow_1 \uparrow_2 \uparrow_3\rangle -2 |\uparrow_1 \downarrow_2 \uparrow_3\rangle \right] \nonumber\\
|3\rangle &=& (201) = |S\rangle_{11} |\uparrow\rangle_3 = |\uparrow_1 \downarrow_1 ; \uparrow_3 \rangle \label{eq:basis6}\\
|4\rangle &=& (102) = |\uparrow\rangle_1 |S\rangle_{33} = |\uparrow_1  ; \uparrow_3 \downarrow_3 \rangle \nonumber\\
|5\rangle &=& (120) = |\uparrow\rangle_1 |S\rangle_{22} = |\uparrow_1  ; \uparrow_2 \downarrow_2 \rangle \nonumber\\
|6\rangle &=& (021) = |S\rangle_{22} |\uparrow\rangle_3 = |\uparrow_2 \downarrow_2 ; \uparrow_3 \rangle \nonumber ~. 
\end{eqnarray}  
and the Hamiltonian matrix is
\begin{widetext}
\begin{equation}\label{eq:Hmatrix}
H = E_0 +  \left( 
\begin{array}{cccccc}
0 & 0 & \frac{t_l}{\sqrt{2}} & \frac{t_r}{\sqrt{2}} & \frac{t_r}{\sqrt{2}} & \frac{t_l}{\sqrt{2}}  \\
0 & 0 & \frac{\sqrt{6}}{2}t_l & -\frac{\sqrt{6}}{2}t_r & -\frac{\sqrt{6}}{2}t_r & \frac{\sqrt{6}}{2}t_l \\
\frac{t_l}{\sqrt{2}} & \frac{\sqrt{6}}{2}t_l & \varepsilon-\varepsilon_M+\widetilde{U}_1 & 0 & 0 & 0 \\
\frac{t_r}{\sqrt{2}} & -\frac{\sqrt{6}}{2}t_r & 0 & -\varepsilon-\varepsilon_M+\widetilde{U}_3 & 0 & 0 \\ 
\frac{t_r}{\sqrt{2}} & -\frac{\sqrt{6}}{2}t_r & 0 & 0 & \varepsilon+\varepsilon_M+\widetilde{U}_2 & 0 \\ 
\frac{t_l}{\sqrt{2}} & \frac{\sqrt{6}}{2}t_l & 0 & 0 & 0 & -\varepsilon+\varepsilon_M+\widetilde{U}'_2
\end{array}
\right) ~,
\end{equation}
\end{widetext}
where $E_0$=$3\bar{\vep} + V_{12} + V_{23} + V_{13}$, $\widetilde{U}_1$=$U_1 - V_{12} - V_{23} + V_{13}$, $\widetilde{U}_2$=$U_2 + V_{12} - V_{23} - V_{13}$, $\widetilde{U}'_2$=$U_2 - V_{12} + V_{23} - V_{13}$, and $\widetilde{U}_3$=$U_3 - V_{12} - V_{23} + V_{13}$. 
$\widetilde{U}_i$'s are the Coulomb interaction energy changes when $i$-th dot is doubly occupied by tunneling of an electron from an adjacent dot. $\widetilde{U}_2$ is for tunneling from the right dot, i.e. (120) configuration, and $\widetilde{U}'_2$ is for tunneling from the left dot, i.e. (021) configuration.

We assume $U_i \gg V_{ij} \gg t_l,t_r$ which is true for typical QD devices and that the TQD is in resonant TQD regime where $|\vep|, |\vep_M| \ll U_i$.
Then we can perform Schrieffer-Wolff transformation \cite{SW_transform_pr1966} to obtain an effective Hamiltonian 
in a subspace spanned by two qubit states $\{ |0\rangle_Q=|2\rangle, |1\rangle_Q=|1\rangle\}$.
\begin{equation}\label{eq:Heff}
\hat{H}_{\mathrm{eff}} = \left(E_0 -\frac{J_l+J_r}{2} \right) \openone - \frac{J}{2} \hat{\sigma}_z - \frac{\sqrt{3}j}{2} \hat{\sigma}_x~,
\end{equation}
where $J$=$(J_l+J_r)/2$, and $j$=$(J_l-J_r)/2$. 
The exchange interactions between dots 1 and 2 ($J_l$) and dots 2 and 3 ($J_r$) were defined as 
$J_l$=$2 t_l^2 (\widetilde{U}_1 + \widetilde{U}'_2)/f_l(\vep,\vep_M)$ and $J_r$=$2 t_r^2 (\widetilde{U}_2 + \widetilde{U}_3)/f_r(\vep,\vep_M)$ 
where $f_l(\vep,\vep_M)$=$\widetilde{U}_1 \widetilde{U}'_2 - ( \widetilde{U}_1 - \widetilde{U}'_2 ) (\vep-\vep_M) - (\vep-\vep_M)^2$ and 
$f_r(\vep,\vep_M)$=$\widetilde{U}_2 \widetilde{U}_3 - ( \widetilde{U}_2 - \widetilde{U}_3 ) (\vep+\vep_M) - (\vep+\vep_M)^2$.

Note that Eq. (\ref{eq:Heff}) is equivalent to exchange-coupled Heisenberg Hamiltonian $\hat{H}_{\mathrm{eff}}$=$E_0\openone + J_l \hat{\mathbf{s}_1}\cdot\hat{\mathbf{s}_2} + J_r \hat{\mathbf{s}_2}\cdot\hat{\mathbf{s}_3}$ in the qubit space where $\hat{\mathbf{s}_j}$ is the spin operator for the electron spin in $j$-th QD.


{\it Sweet spot and single qubit operations--} 
The sweet spots are defined where the energy gap of the effective Hamiltonian in the qubit space [Eq. (\ref{eq:Heff})] is insensitive to the charge noise in detuning parameters up to first order.
That is, where $E_{01}$=$\sqrt{J^2+3j^2}$=$\sqrt{J_l^2+J_r^2-J_l J_r}$ is immune to small variations in the parameters $\vep$ and $\vep_M$.
From $\partial E_{01} /\partial \vep$=$\partial E_{01} /\partial \vep_M$=0, we obtain the sweet spot
\begin{eqnarray}
\vep &=& \frac{1}{4}\left( -\widetilde{U}_1 + \widetilde{U}'_2 - \widetilde{U}_2 + \widetilde{U}_3\right) \label{eq:sweet_spot_ep}\\
\vep_M &=& \frac{1}{4}\left( \widetilde{U}_1 - \widetilde{U}'_2 - \widetilde{U}_2 + \widetilde{U}_3\right)\label{eq:sweet_spot_epM}~.
\end{eqnarray}
Alternatively, the sweet spot can be obtained from the conditions
$\partial J_l/\partial\vep$=$\partial J_l/\partial\vep_M$=$\partial J_r/\partial\vep$=$\partial J_r/\partial\vep_M$=0, since the energy gap $E_{01}$ is a function of $J_l$ and $J_r$. 
Note that this sweet spot does not depend on the tunneling $t_l$ and $t_r$, allowing tunabilty for $J_l$ and $J_r$ while staying on the sweet spot.
At the sweet spot, 
$J_l$=$8 t_l^2/(\widetilde{U}_1 + \widetilde{U}'_2)$ and $J_r$=$8 t_r^2/(\widetilde{U}_2 + \widetilde{U}_3)$
and we can implement single qubit operations by tuning $t_l$ and $t_r$.

\begin{figure}
  \includegraphics[width=\linewidth]{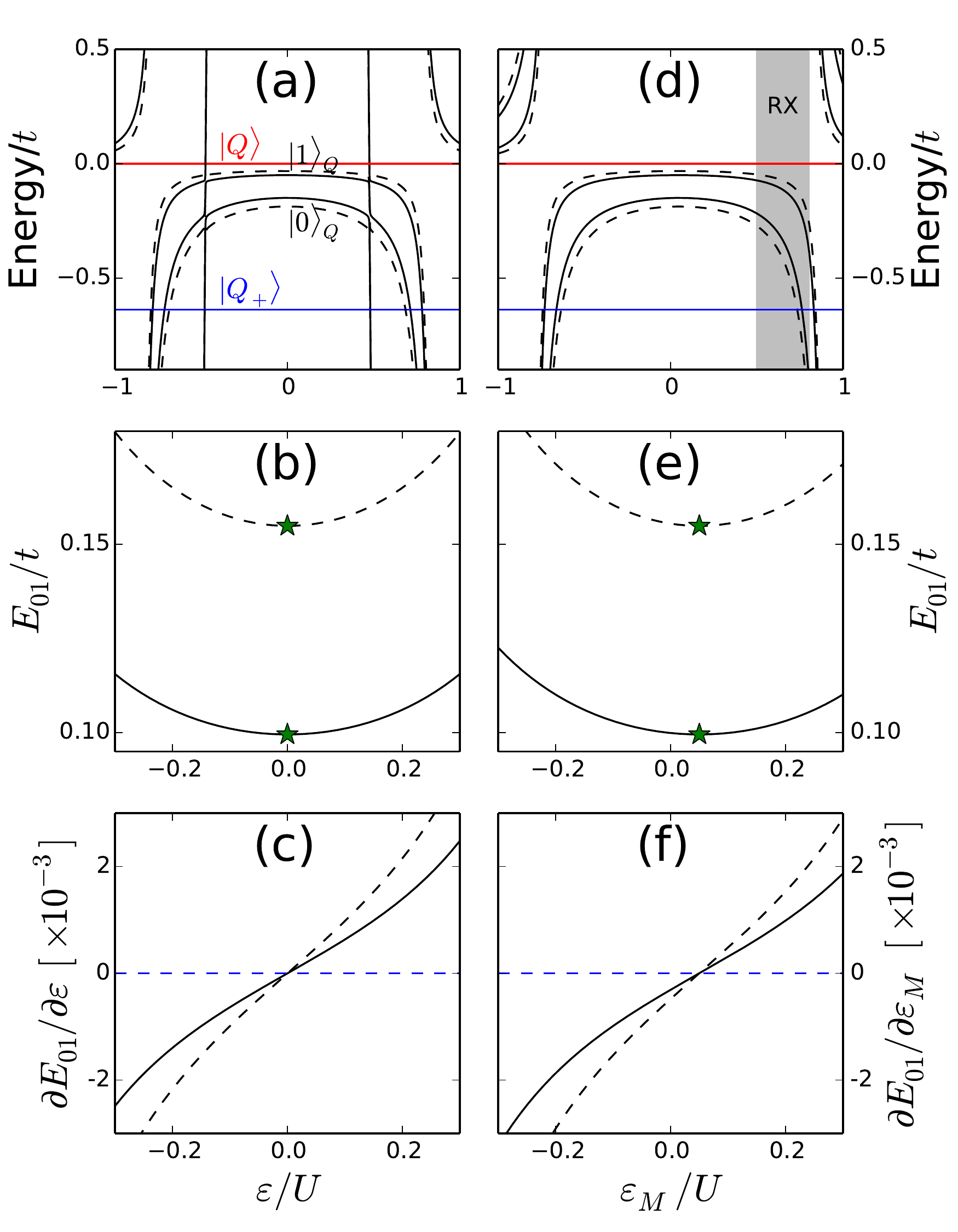}\\
  \caption{Always-on exchange-only qubit (AEON) with $t_l$=$t_r$=$t$ (solid curves) and $t_l$=$(1+\sqrt{3}-\sqrt{2})t$ and $t_r$=$(1-\sqrt{3}+\sqrt{2})t$ (dashed curves) with $t$=0.02$U$.
  (a) Energy spectrum of the TQD device as a function of $\vep$ with fixed $\vep_M$=0.05$U$. The sweet spots indicated with green stars in (b) both correspond to $\vep$=0. Similarly, (d) shows the energy spectrum as a function of $\vep_M$ with fixed $\vep$=0. The sweet spots in (e) both correspond to $\vep_M$=0.05$U$ as was shown in (f). The blue dashed line in (c) and (f) are guides for zero derivative. The independence of the sweet spon on $t_l$ and $t_r$ is crucial to implementation of full set of logical gates on the sweet spot.}
  \label{fig:result}
\end{figure}

Figure \ref{fig:result} shows numerical results by exactly solving the Hubbard Hamiltonian, Eq. (\ref{eq:Hubbard}), also including (210) and (012) configurations. We used parameters $U_1$=$U_3$=$U$=1meV, $U_2$=0.8$U$, $V_{12}$=$V_{23}$=0.1$U$, $V_{13}$=0.05$U$. The Zeeman energy was obtained using GaAs material parameters with $B_{\mathrm{ext}}$=0.5T. Note that the results of this work would apply equally well to silicon quantum dots or other possible semiconductor systems. With these parameters the sweet spot is $\vep$=0 and $\vep_M$=0.05$U$.
In addition to the states with $S_{\mathrm{tot}}$=$S_{\mathrm{tot}}^z$=1/2 that comprise the subspace, there are two more low-energy states $|Q\rangle$=$\left(|\uparrow_1\uparrow_2\downarrow_3\rangle + |\uparrow_1\downarrow_2\uparrow_3\rangle + |\downarrow_1\uparrow_2\uparrow_3\rangle  \right)/\sqrt{3}$ and $|Q_+\rangle$=$|\uparrow_1\uparrow_2\uparrow_3\rangle$. But these states do not interact with any states in the subspace and can be neglected.
We calculated the energy spectrum and the sweet spot for $t_l$=$t_r$ (solid curves) corresponding to a rotation around $\hat{\mathbf{z}}$ axis, and $t_l$=$(\sqrt{6}+\sqrt{2})t_r/2$ (dashed curves) corresponding to a rotation around $\hat{\mathbf{n}}$=$-(\hat{\mathbf{x}}+\hat{\mathbf{z}})/\sqrt{2}$ axis which, in combination with the Pauli $Z$ gate, can be used to implement Pauli $X$ gate \cite{Hanson_Burkard_prl2007,TQD_sweet_spot_Fei_Friesen_prb2015}, $X$=$R_{\hat{\bf{n}}}(\pi) Z R_{\hat{\bf{n}}}(\pi)$ where $R_{\hat{\bf{n}}}(\pi)$ is a $\pi$ rotation around $\hat{\mathbf{n}}$ and $Z$ is Pauli $Z$ gate which is a $\pi$ rotation around $\hat{\mathbf{z}}$ axis.   
Figure \ref{fig:result}(a) shows the energy spectrum as a function of $\vep$ with fixed $\vep_M$=0.05$U$. The energy difference between the two qubit states $|0\rangle_Q$ and $|1\rangle_Q$ shows minimum at the sweet spot $\vep$=0, marked with green stars [Fig. \ref{fig:result}(b)] and the derivative vanishes there [Fig. \ref{fig:result}(c)].  
As a function of $\vep_M$, the energy spectrum at fixed $\vep$=0 is in Fig. \ref{fig:result}(d) and the qubit energy [Fig. \ref{fig:result}(e)] and its derivative [Fig. \ref{fig:result}(f)] shows that the sweet spot $\vep_M$=0.05$U$ does not change for different $t_l$ and $t_r$.


\begin{figure}
  \includegraphics[width=\linewidth]{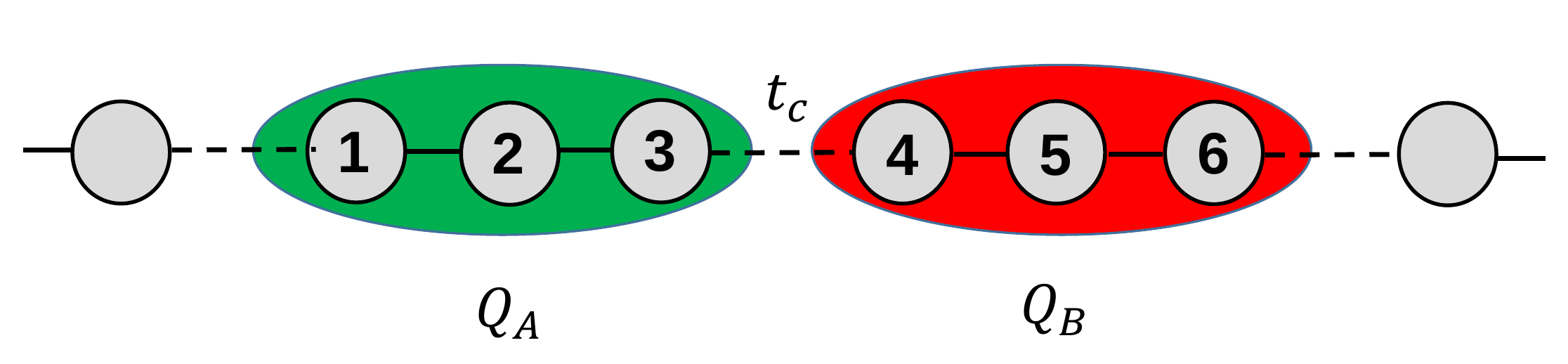}\\
  \caption{An array of always-on, exchange-only qubits as described in this paper.  The sweet spot of each encoded qubit changes due to the inter-qubit Coulomb interactions, but still all gate operations can be done on a sweet spot. The qubit-qubit interaction is implemented using the exchange coupling between neighboring dots belonging to different encoded qubits, which can also be made to be insensitive to charge noise up to first order by tuning the qubit energies.}
  \label{fig:array}
\end{figure}

{\it Two-qubit operations--} 
Two-qubit gates between exchange-only qubits require a long sequence of pair-wise exchange interaction pulses  \cite{divincenzo_bacon_nature2000,fong_wandzura_qic2011}, if the pulses are applied in a serial mode. Keeping the intra-qubit exchange interactions always-on allows for a much shorter sequence as was shown in \cite{RX_2q_gate_Doherty_prl2013}. The procedure described in the previous section allows individual control of the intra-qubit exchange interactions while staying on a sweet spot, and now we show that the {\em inter}-qubit exchange interaction can also be insensitive to the charge noise up to first order. 

Let's consider an array of exchange-only qubits as depicted in Fig. \ref{fig:array}. First, the position of the sweet spot in the parameter space $(\vep,\vep_M)$ shifts because we need to redefine the Coulomb interaction change $\widetilde{U}_i$ due to double occupancy to take into account the presence of additional electron spins in different qubits. For example, $\widetilde{U}_1$=$U_1 - V_{12} - V_{23} + V_{13} + V_{\Sigma,1} - V_{\Sigma,2}$ where $V_{\Sigma,i}$ is sum of $V_{ij}$ for all QD $j$ belonging to a different qubit (e.g. $V_{\Sigma,1}$=$V_{14}+V_{15}+V_{16}+\ldots$).  Similarly, $\widetilde{U}_3$=$U_3 - V_{12} - V_{23} + V_{13} + V_{\Sigma,3} - V_{\Sigma,2}$, $\widetilde{U}_2$=$U_2 + V_{12} - V_{23} - V_{13} + V_{\Sigma,2} - V_{\Sigma,3}$, and $\widetilde{U}'_2$=$U_2 - V_{12} + V_{23} - V_{13} + V_{\Sigma,2} - V_{\Sigma,1}$. The sweet spot condition in Eqs. (\ref{eq:sweet_spot_ep}) and (\ref{eq:sweet_spot_epM}) remains the same with the above adjusted $\widetilde{U}_i$'s. 

The inter-qubit exchange coupling $J_c$ between QDs 3 and 4 is given by
\begin{equation}
J_c = 2 t_c^2 \left[ \frac{1}{\overline{U}_3 + \vep_3 - \vep_4} + \frac{1}{\overline{U}_4 - \vep_3 + \vep_4} \right] ~,
\end{equation}
where $\overline{U}_3$=$U_3+V_{13}+V_{23}-V_{34}-V_{45}-V_{46}+V_{\Sigma,3}-V_{\Sigma,4}$ and $\overline{U}_4$=$U_4+V_{45}+V_{46}-V_{13}-V_{23}-V_{34}+V_{\Sigma,4}-V_{\Sigma,3}$.
To find the sweet spot for $J_c$, from $\partial J_c/\partial\vep_3$=$\partial J_c/\partial\vep_4$=0, we obtain
$\vep_3-\vep_4$=$-(\overline{U}_3-\overline{U}_4)/2$. This can be satisfied by tuning $\bar{\vep}^{(A)}$ and  $\bar{\vep}^{(B)}$, while $\vep^{(A)}$,$\vep_M^{(A)}$,$\vep^{(B)}$,$\vep_M^{(B)}$ remain at the sweet spots for qubits $A$ and $B$.
Similar to the intra-qubit exchange interactions, $J_c$ is controlled by tuning the tunneling $t_c$ which does not affect the sweet spot.

In the weak coupling regime where $J_c \ll J^{(A)}, J^{(B)}$, this exchange coupling leads to a coupling Hamiltonian \cite{RX_2q_gate_Doherty_prl2013}
\begin{eqnarray}\label{eq:Hc}
\widehat{H}_c &=& \delta J_z \left( \sigma_{zA} + \sigma_{zB} \right) /2 + J_{zz} \sigma_{zA} \sigma_{zB} \nonumber\\
              &&  + J_{\perp} \left( \sigma_{xA} \sigma_{xB} + \sigma_{yA} \sigma_{yB} \right) ~,
\end{eqnarray}
where $\sigma_{\alpha A}$ ($\sigma_{\alpha B}$) ($\alpha$=$x,y,z$) is the Pauli operator for qubit $A$ ($B$).
The coupling coefficients $\delta J_z$, $J_{zz}$, and $J_{\perp}$ are all proportional to the exchange coupling $J_c$.
For the linear geometry in Fig. \ref{fig:array}, $\delta J_z/J_c$=$J_{zz}/J_c$=1/36. $J_{\perp}/J_c$=-1/24 for $J^{(A)} \simeq J^{(B)}$ and 0 for $|J^{(A)}-J^{(B)}| \gg J_c$ \cite{RX_2q_gate_Doherty_prl2013}.

We can estimate the gate time for a CPHASE gate using this approach. Implementation of CPHASE gate is simpler for $J_{\perp}/J_c$=0 (i.e., $|J^{(A)}-J^{(B)}| \gg J_c$). For typical exchange coupling strength of hundreds of MHz for  $J^{(A)}/h$ and $J^{(B)}/h$ for current QD devices \cite{Shulman_Yacoby_science2012,DQD_sweet_spot_Reed_Hunter_2015}, for example, $J^{(A)}/h$=100 MHz and $J^{(B)}/h$=300 MHz, we can use $J_c \simeq$ 10MHz, and the CZ gate can be obtained for $\int J_{zz}(t) dt$=$\pi/4$ which corresponds to a gate operation time of a few hundred ns for a square pulse. This is somewhat slow compared to the typical single exchange pulse of a few ns, due to the requirement of small $J_c$ to prevent leakage errors, but a different geometry for qubit-qubit coupling could help reduce the operation time \cite{RX_2q_gate_Doherty_prl2013}. For example, if we swap spins in QDs 2 and 3 and also swap spins in QDs 4 and 5, the inter-qubit exchange coupling is equivalent to the "butterfly" geometry in Ref. \cite{RX_2q_gate_Doherty_prl2013} and the two-qubit CZ gate can be done in about 20ns.



\begin{table*}
\begin{ruledtabular}
\begin{tabular}{llll}
Qubit type & 3-DFS exchange-only \cite{divincenzo_bacon_nature2000, fong_wandzura_qic2011} & RX \cite{RX_theory_Taylor_prl2013,RX_exp_Medford_prl2013,RX_2q_gate_Doherty_prl2013} & AEON \\
 \hline
QD levels & General & $\vep_2 \gg \vep_1 \simeq \vep_3$ & $\vep_1 \simeq \vep_2 \simeq \vep_3$ \\
Coherence protection & DFS & DFS, partial sweet spot & DFS, full sweet spot \\
Idle/memory  & all exchange couplings off & always-on coupling, $f_Q \sim$ 0.5-2GHz & all exchange couplings off or always-on \\
Single qubit gates & 4 fast pair-wise pulses & RF pulse & 3 fast simultaneous pair-wise pulses \\
Two-qubit gates & 18 fast pair-wise pulses & dipole-dipole or single exchange pulse & single exchange pulse 
\end{tabular}
\end{ruledtabular}
\caption{\label{tab:comparison} Comparison between different types of exchange-only qubits.} 
\end{table*}

{\it Comparison and conversion to other encoded 3-spin qubits--} 
For the RX qubit \cite{RX_theory_Taylor_prl2013,RX_exp_Medford_prl2013}, the TQD device is tuned to be in a regime where the (111) configuration is close to (201) and (102) configurations for initialization and readout. This implies that $\vep_M$ is relatively large, comparable to the on-site Coulomb interaction $U$. The RX regime is shaded with gray in Fig. \ref{fig:result}(d). The sweet spot of the RX qubit is therefore only insensitive to $\vep$ fluctuations and sensitive to $\vep_M$ variations in the first order. A full sweet spot where the qubit frequency is insensitive to both $\vep$ and $\vep_M$ is only possible outside of (111) configuration \cite{TQD_sweet_spot_Burkard_prb2015}, which limits the qubit performance.
Furthermore, the sweet spot in the RX qubit depends on the asymmetry of the tunneling elements $t_l$ and $t_r$, requiring a different method (i.e. microwave control) to implement full single qubit rotations. The logical single-qubit gate operations are implemented using microwave at the partial sweet spot determined by $t_l$ and $t_r$.

In comparison, the sweet spot we suggest here in Eqs. (\ref{eq:sweet_spot_ep}) and (\ref{eq:sweet_spot_epM}) for an AEON qubit is more general. It does not require any symmetries in parameters, and it does not depend on changing $t_l$ and $t_r$, allowing for full logical gate operations on the sweet spot by tuning the tunnel coupling between dots, i.e., all-DC control. It is in a deep (111) regime where $\vep_i$'s are of similar values compared to the Coulomb interaction energy $U_i$'s (resonant TQD regime). In this regime, increasing $\vep$ leads to (012) ground state, not (102) ground state as in RX regime. (The two almost vertical lines in Fig. \ref{fig:result}(a) correspond to states with (012) and (210) configurations.) This can be a problem for initialization or readout, since (012) is not coupled to (111) by a single tunneling event and the anti-crossing between them is very small, that is, a long time is needed to move from (012) to (111); but this is beneficial for turning off any dipole-like coupling to external noise or quantum systems. One way to do the initialization/readout for an AEON qubit would be to initialize in the RX regime and then move to the sweet spot in the resonant TQD regime, by tuning $\vep_M$. Readout can be done in a reverse order moving from resonant TQD regime to RX regime. Alternatively, converting to the RX qubit can turn on coupling to a resonator for beyond-nearest neighbor quantum gates.

The original exchange-only qubit \cite{divincenzo_bacon_nature2000} was defined as the singlet/triplet states of QDs 1 and 2, while we defined our qubit states as the singlet/triplet states between QDs 1 and 3 [see Eq. (\ref{eq:basis6})]. The sweet spot defined in Eqs. (\ref{eq:sweet_spot_ep}) and (\ref{eq:sweet_spot_epM}) is also a sweet spot for the original exchange-only qubit and all exchange operations can also be realized on the sweet spot by tuning the tunneling $t_l$ and $t_r$. 
Conventionally a serial sequence of exchange pulses were proposed to implement the logical gate operations for the original exchange-only qubit, and rather long sequences of exchange operations are required for two-qubit gates \cite{divincenzo_bacon_nature2000,fong_wandzura_qic2011}. Simultaneous exchange operations (``always-on") can reduce the length of the sequence \cite{two_step_sq_gate,RX_2q_gate_Doherty_prl2013} and we show here that this is possible all on a sweet spot. 
Sometimes it is beneficial to work on one or the other definition for the qubit states, and they can be easily converted to and from each other by applying a swap operation between QDs 2 and 3 which can be implemented by tuning $J_l$=0 and $J_r \neq$ 0. 

Comparison between different types of exchange-only qubits is given in Table \ref{tab:comparison}.


In summary, it was shown that there exists a sweet spot for an always-on, exchange-only qubit in a linear array of three quantum dots where the qubit energy is only second order sensitive to charge noise for both critical system parameters representing changes in the dot energy levels, and that all single- and two-qubit encoded gates can be implemented on this sweet spot. 

This work shows that multiple exchange interaction gates can be applied simultaneously while remaining on the sweet spot in exchange-only qubits. It should also be applicable in coupling different types of QD spin encoded qubits \cite{Mehl_DiVincenzo_prb2015} since the exchange interaction enables coupling between them, or in other spin qubit systems such as impurity spins \cite{kane_nature1998} which offer tunable detuning and tunnel barriers.  

Another advantage of the sweet spot operations studied here is that this allows for a true off-state for the encoded qubit by turning off the exchange interactions (transforming the qubit into the traditional 3-DFS qubit), while remaining on the sweet spot; this simplifies gate operations and can be useful for storing quantum information.



\bibliographystyle{apsrev4-1}

%

\end{document}